\documentclass[12pt,thmsa,a4paper]{article}
\usepackage{sw20aip}

\input tcilatex
\QQQ{Language}{
American English
}

\begin{document}

\author{Hans J. Wospakrik\thanks{%
email: hansjw@fi.itb.ac.id} \\
\\
Theoretical Physics Laboratory, \\
Department of Physics, Institute of Technology Bandung,\\
Jalan Ganesha 10, Bandung 40132, Indonesia}
\title{Path Integral Evaluation of the Free Propagator on the ($D$-1)-dimensional
Pseudosphere}
\date{}
\maketitle

\begin{abstract}
We present an explicit path integral evaluation of the free Hamiltonian
propagator on the $(D-1)$-dimensional pseudosphere, in the horicyclic
coordinates, using the integral equation method. This method consists in
deriving an integral equation for the propagator that turns out to be of
Abel's type.
\end{abstract}

\section{Introduction}

Recently, significant advances had been achieved in the formulation of the
path integral on the hyperbolic space, especially the pseudosphere$^1$. For
the 2-dimensional pseudosphere $\Lambda ^{(2)}$, the explicit path integral
evaluation of the propagator (heat kernel) $K$ of the free Hamiltonian
(Laplace-Beltrami) operator was first evaluated by Grosche and Steiner$^2$,
and Kubo$^3$ using different methods. Grosche then evaluated the 3- and $(D$%
-1$)$-dimensional cases, in the horicyclic, spherical, and equidistant
coordinates$^{1,4}$.

In this work, we will also consider the same path integral problem, however
we will be using a different method which is basically an extension of
Kubo's method in the 2-dimensional case, where we will restrict ourself to
the horicyclic coordinates. In applying this method, we integrate out first
the path integral solution of the $(D$-2$)$-dimensional ''Euclidean space''
coordinates which then results in an Abel's type integral equation for $K$
that we then solve explicitly.

\section{Pseudosphere and Path Integral}

Let us begin with recalling the basic facts pertaining to the
parametrization of the ($D-1)$ pseudosphere $\Lambda ^{(D-1)}$ in the
horicyclic coordinates and the path integral formulation of the propagator
of the free Hamiltonian operator.

The pseudosphere $\Lambda ^{(D-1)}$ is defined as one sheet of the $(D$-1$)$%
-dimensional double sheeted hyperboloid:

\begin{equation}
(Z^0)^2-(Z^1)^2-(Z^2)^2-...-(Z^{D-1})^2=1,  \tag{2.1}
\end{equation}
with $Z^0>0$, in the $D$-dimensional hyperbolic space with the pseudometric:

\begin{equation}
ds^2=(dZ^0)^2-(dZ^1)^2-(dZ^2)^2-...-(dZ^{D-1})^2,  \tag{2.2}
\end{equation}
where the group $SO(D$-1,1$)$ acts transitively, since $\Lambda
^{(D-1)}=SO(D $-1$,1)/SO(D$-1$)^1.$

Introducing the horicyclic coordinates,

\begin{equation}
y=1/(Z^0+Z^1),\text{ }x^\mu =yZ^{\mu +1},\text{ }(\mu =1,2,...,D\text{-2}), 
\tag{2.3}
\end{equation}
the metric (2.2) becomes

\begin{equation}
ds^2=\frac{[(dy)^2+\sum_{\mu =1}^{D-2}(dx^\mu )^2]}{y^2},  \tag{2.4}
\end{equation}
(up to an overall minus sign factor). This is nothing but the $(D$-1$)$%
-dimensional Poincare upper half ''plane'' or space ($y>0)$ representation
for $\Lambda ^{(D-1)}$.

Let ($q^0=y,q^\mu =x^\mu )=(q^a)=$ \textbf{q}, then the metric tensor
components of the metric (2.4) read,

\begin{equation}
g_{ab}=y^{-2}\delta _{ab},  \tag{2.5}
\end{equation}
where $\delta _{ab}$ ($a,b=0,1,...,D$-2$)$ is the Kronecker's delta.

The free Hamiltonian operator on $\Lambda ^{(D-1)}$ is,

\begin{equation}
\text{\textbf{H}}=-\frac{\hbar ^2}{2m}\Delta _{LB},  \tag{2.6$a$}
\end{equation}
where $\Delta _{LB}$ is the Laplace-Beltrami operator,

\begin{eqnarray}
\Delta _{LB} &=&\frac 1{\sqrt{\det (g_{ab})}}\frac \partial {\partial
q^c}\left( \sqrt{\det (g_{ab})}g^{cd}\frac \partial {\partial q^d}\right) 
\tag{2.6$b$} \\
&=&y^2\left( \frac{\partial ^2}{\partial y^2}+\frac{\partial ^2}{\partial
(x^1)^2}+\frac{\partial ^2}{\partial (x^2)^2}+...+\frac{\partial ^2}{%
\partial (x^{D-2})^2}\right) -(D-3)y\frac \partial {\partial y}  \nonumber
\end{eqnarray}

The path-integral representation of the propagator of the free Hamiltonian
operator (2.6) is given by$^1$:

\begin{equation}
K(\mathbf{q}^{\prime \prime },t^{\prime \prime },\mathbf{q}^{\prime
},t^{\prime })=\dint \left[ \frac{dy}y\right] \dprod_{\mu =1}^{D-2}\left[ 
\frac{dx^\mu }y\right] \exp [\frac i\hbar S],  \tag{2.7$a$}
\end{equation}
where $S$ is the action: 
\begin{equation}
S=\int\limits_{t^{\prime }}^{t^{\prime \prime }}dt\left\{ (\frac m2)(\frac
1{y^2})\left[ \left( \frac{dy}{dt}\right) ^2+\sum_{\mu =1}^{D-2}\left( \frac{%
dx^\mu }{dt}\right) ^2-\frac{\hbar ^2}{4m^2}(D-1)(D-3)y^2\right] \right\} . 
\tag{2.7$b$}
\end{equation}

The integration over the $(D$-2$)$-dimensional (path) coordinates ($x^\mu )$%
\textbf{\ }in (2.7) are carried out explicitly in the following symmetrical
discretized form$^{1,3}$, 
\[
\dint \left[ \frac{dx_\mu }y\right] \exp \left\{ \frac{im}{2\hbar }\right\}
\int\limits_{t^{\prime }}^{t^{\prime \prime }}dt\left\{ \frac 1{y^2}\left( 
\frac{dx^\mu }{dt}\right) ^2\right\} = 
\]

\[
\lim_{N\rightarrow \infty }\left( \frac \alpha {\pi \varepsilon }\right)
^{N/2}\prod_{r=1}^{N-1}\int\limits_{-\infty }^\infty \left( \frac{dx_r^\mu }{%
y_r}\right) \exp \left\{ -\alpha \sum_{n=0}^{N-1}\frac{(x_{_{n+1}}^\mu
-x_{_n}^\mu )^2}{\varepsilon y_{n+1}y_n}\right\} = 
\]

\begin{equation}
(y^{\prime \prime }y^{\prime })^{1/2}\left( \frac \alpha \pi \right)
^{1/2}Y^{-1/2}\exp \left\{ -\alpha \frac{(x^{^{\prime \prime }\mu
}-x^{\prime \mu })^2}Y\right\} ,  \tag{2.8$a$}
\end{equation}
where $\varepsilon =(t_{n+1}-t_n)=(t"-t^{\prime })/N,$ $\alpha =(m/2i\hbar )$%
, and based on the above discretized approximation, the functional $Y$ is
given by, 
\begin{equation}
Y=\lim_{N\rightarrow \infty }\sum_{n=0}^{N-1}\varepsilon \text{ }%
y_ny_{n+1}=\int\limits_{t^{\prime }}^{t^{\prime \prime }}dt\text{ }y^2 
\tag{2.8$b$}
\end{equation}
Using the path integral solution (2.8) in (2.7) gives

\[
K(\mathbf{q}^{\prime \prime },t^{\prime \prime },\mathbf{q}^{\prime
},t^{\prime })=\left( \frac \alpha \pi \right) ^{(D-2)/2}(y^{\prime \prime
}y^{\prime })^{(D-2)/2}\dint \left[ \frac{dy}y\right] Y^{(2-D)/2} 
\]
\begin{equation}
\exp \left[ -\alpha \frac{R^2}Y-\alpha \int\limits_{t^{\prime }}^{t^{\prime
\prime }}dt\left\{ \frac 1{y^2}\left( \frac{dy}{dt}\right) ^2-\beta \right\}
\right] ,  \tag{2.9}
\end{equation}
where $R^2=\sum_{\mu =1}^{D-2}\xi ^\mu \xi ^\mu ,$ with $\xi ^\mu
=(x^{^{\prime \prime }\mu }-x^{\prime \mu }),$ and $\beta =(\hbar
^2/4m^2)(D-1)(D-3).$

Thus, the propagator $K$ is a function of $R^2.$ To get rid of the $Y$%
-factor in evaluating the $y$-integral in (2.9), let us integrate out the
path integral (2.9) symmetrically with respect to the ($\xi ^\mu )$
coordinates, which results in the following simple relation:

\[
\int\limits_{-\infty }^\infty \int\limits_{-\infty }^\infty
...\int\limits_{-\infty }^\infty d\xi ^1d\xi ^2...d\xi
^{(D-2)}K(R^2,y^{\prime \prime },t^{\prime \prime },y^{\prime },t^{\prime
})= 
\]

\begin{equation}
(y^{\prime \prime }y^{\prime })^{(D-2)/2}\dint \left[ \frac{dy}y\right] \exp
\left[ -\alpha \int\limits_{t^{\prime }}^{t^{\prime \prime }}dt\left\{ \frac
1{y^2}\left( \frac{dy}{dt}\right) ^2-\beta \right\} \right] .  \tag{2.10}
\end{equation}

Introducing the new variable: 
\begin{equation}
z=\ln y,  \tag{2.11}
\end{equation}
into the path integral in the right hand side of (2.10), we immediately find
that it is the propagator of a free Euclidean particle$^5$. Let us proceed
by introducing the polar coordinates in the left hand side of (2.10), $i.e.,$
\begin{equation}
(\xi ^1,\xi ^2,...,\xi ^{(D-2)})\rightarrow (R,\phi ,\theta _1,\theta
_2,...,\theta _{(D-4)}),  \tag{2.12$a$}
\end{equation}
where the variables assuming the values:

\begin{equation}
0\leq R<\infty \text{, }0\leq \phi <2\pi ,\text{ }0\leq \theta _r<\pi ,\text{
}(r=1,2,...,D-4).  \tag{2.12$b$}
\end{equation}
By substituting the corresponding free particle propagator solution in the
right hand side of (2.10), and evaluating the angular variables integral, we
obtain the following $integral$ $equation$ for the propagator $K$:

\[
\int\limits_l^\infty dk\text{ }K(k,T)(k-l)^{(D-4)/2}= 
\]

\begin{equation}
\Gamma \left( \frac{D-2}2\right) \left( \frac 1{2\pi }\right)
^{(D-2)/2}\left( \frac \alpha {\pi T}\right) ^{1/2}\exp \left\{ -(\alpha
/T)s^2+\alpha \beta T\right\} ,  \tag{2.13$a$}
\end{equation}
where $T=(t^{^{\prime \prime }}-t^{\prime })$, $\Gamma (p)$ is the Euler
gamma function, and

\begin{equation}
s=\cosh ^{-1}(l),\text{ }l=\frac{y^{\prime \prime 2}+y^{\prime 2}}{%
2y^{\prime \prime }y^{\prime }},\text{ }k=\frac{R^2}{2y^{\prime \prime
}y^{\prime }}+l.  \tag{2.13$b$}
\end{equation}

Thus $\cosh ^{-1}(k)$ is the hyperbolic distance between two $arbitrary$
points on the pseudosphere $\Lambda ^{(D-1)}$ with fixed $y^{\prime }$ and $%
y^{\prime \prime }.$ Note that, since $k$ is a variable of integration, it
is not necessary related to the $fixed$ hyperbolic distance between the
initial and final coordinates $(\mathbf{q}^{\prime },\mathbf{q}^{\prime
\prime })$. Let this fixed hyperbolic distance be denoted by,

\begin{equation}
\text{ }d=d_{\Lambda ^{(D-1)}}(\mathbf{q}^{\prime },\mathbf{q}^{\prime
\prime }).  \tag{2.14}
\end{equation}
Then, since we can always find a transformation which transforms two points $%
\mathbf{q}^{\prime }$ and $\mathbf{q}^{\prime \prime }$ into two points on
the $y$-axis preserving the hyperbolic distance between two points
invariant, so the parameter $s$ is just the invariant distance between $%
\mathbf{q}^{\prime }$ and $\mathbf{q}^{\prime \prime }.$

\section{Solutions for the D = 3, 4, and \TEXTsymbol{>} 4 cases}

In this section, we will be considering the solution of the integral
equation (2.13) for the $D$ = 3, 4, and \TEXTsymbol{>} 4 cases, separately.

\subsection{\textit{The D = 3 case}}

In this case, the corresponding integral equation (2.13) reduces to Kubo's
integral equation$^3$,

\begin{equation}
\int\limits_l^\infty dk\frac{K(k,T)}{\sqrt{k-l}}=\left( \frac \alpha {2\pi
T}\right) ^{1/2}\exp \left\{ -(\alpha /T)s^2+\alpha \beta T\right\} , 
\tag{3.1}
\end{equation}
where in Ref. 3, the solution was given by using the Helgason's method$^7.$
(Note that, for this case $\beta =0!$ We do not discard $\beta $ in order to
compare with other results$^{2,3}$). In the following, we will present
another method which is motivated by the fact that the integral equation
(3.1) is almost of the Abel's type, for which the method of solution is
well-known$^6$.

First of all, we let $l=\cosh (s)$ to be a variable so that $s$ is not
necessary equal to the invariant distance $d$ in eq.(2.14). Then we take the
derivative of eq. (3.1) with respect to $s=\cosh ^{-1}(l),$ which gives,

\begin{equation}
\frac{\partial F}{\partial s}=-2\alpha \left( \frac \alpha {2\pi T}\right)
^{1/2}s\exp \left\{ -(\alpha /T)s^2+\alpha \beta T\right\} ,  \tag{3.2$a$}
\end{equation}
where

\begin{equation}
F(s)=\int\limits_l^\infty \frac{K(k,T)}{\sqrt{k-l}}dk.  \tag{3.2$b$}
\end{equation}
Next, by multiplying eq. (3.2$a$) with $ds\sqrt{l-u},$ where $u=\cosh (d),$
and integrating out the resulting equation with respect to $s$ from $d$ to $%
\infty ,$ we obtain,

\[
\int\limits_u^\infty \frac{dl}{\sqrt{l-u}}\left[ \int\limits_l^\infty dk%
\frac{K(k,T)}{\sqrt{k-l}}\right] = 
\]
\begin{equation}
\frac{4\alpha }T\left( \frac \alpha {2\pi T}\right)
^{1/2}\int\limits_{d(u)}^\infty ds\text{ }s\sqrt{\cosh (s)-u}\exp \left\{
-(\alpha /T)s^2+\alpha \beta T\right\} ,  \tag{3.3}
\end{equation}
where in the left hand side we have changed the variable of integration from 
$s$ to $l=\cosh (s).$ By reversing the order of integrations with
appropriate changes in the integration limits, then straightforward
manipulations give,

\begin{equation}
\int\limits_u^\infty \frac{dl}{\sqrt{l-u}}\left[ \int\limits_l^\infty dk%
\frac{K(k,T)}{\sqrt{k-l}}\right] =\pi \int\limits_u^\infty dk\text{ }K(k,T).
\tag{3.4}
\end{equation}

Substituting eq. (3.4) into the left hand side of eq. (3.3) and taking the
derivative of the resulting equation with respect to $u=\cosh (d)$, we
finally obtain$^{2,3}$,

\begin{equation}
K(\mathbf{q}^{\prime \prime },t^{\prime \prime },\mathbf{q}^{\prime
},t^{\prime })=\sqrt{2}\left( \frac \alpha {\pi T}\right)
^{3/2}\int\limits_d^\infty ds\text{ }\frac{s\exp \left\{ -(\alpha
/T)s^2+\alpha \beta T\right\} }{\sqrt{\cosh (s)-\cosh (d)}},  \tag{3.5}
\end{equation}
which is nothing but the celebrated McKean's result$^8$ (up to the constants 
$\alpha $ and $\beta ).$

\subsection{\textit{The D = 4 case}}

This is the simplest case, since the integrand factor $(k-l)^{(D-4)}=1$, so
by differentiating eq.(2.13) with respect to $l$, we get the solution,

\begin{equation}
K(\mathbf{q}^{\prime \prime },t^{\prime \prime },\mathbf{q}^{\prime
},t^{\prime })=\left( \frac \alpha {\pi T}\right) ^{3/2}\frac{s\exp \left\{
-(\alpha /T)s^2+\alpha \beta T\right\} }{\sinh (s)}.  \tag{3.6}
\end{equation}
This is the same with the result that was obtained by Grosche$^4.$

\subsection{\textit{The D \TEXTsymbol{>} 4 case}}

It is obvious now that the solution of the integral equation (2.13), for
these cases, could be obtained by taking the derivatives of eq. (2.13) with
respect to $l$; however, the $D$ = $even$, and $odd$ cases should be treated
differently. For the purpose of the derivation, let us express the right
hand side function in (2.13) and its derivatives by,

\begin{equation}
G(s)=\left( \frac \alpha {\pi T}\right) ^{1/2}\exp \left\{ -(\alpha
/T)s^2+\alpha \beta T\right\} \equiv G^{(0)}(s),  \tag{3.7}
\end{equation}

\begin{equation}
G^{(n)}(s)\equiv \frac{\partial ^n}{\partial l^n}G(s(l))=\left[ \frac
1{\sinh (s)}\frac \partial {\partial s}\right] ^nG(s),  \tag{3.8}
\end{equation}
Unfortunately, we have not been able to give a closed expression for $%
G^{(n)}(s).$

\subsubsection{\textit{D = even}}

The solution of the integral equation (2.13) for $K$, is straightforwardly
obtained by taking the derivative with respect to $l$, $(D-2)/2$-times,
which gives the result,

\begin{equation}
K(\mathbf{q}^{\prime \prime },t^{\prime \prime },\mathbf{q}^{\prime
},t^{\prime })=\left( -\frac 1{2\pi }\right) ^{(D-2)/2}G^{(D-2)/2}(s). 
\tag{3.9}
\end{equation}

\subsubsection{\textit{D = odd}}

For these cases, we first take the derivative of (2.13) with respect to $l$, 
$(D-3)/2$-times, which gives the result,

\begin{equation}
\int\limits_l^\infty dk\text{ }\frac{K(k,T)}{\sqrt{k-l}}=(-1)^{(D-3)/2}\sqrt{%
\pi }\left( \frac 1{2\pi }\right) ^{(D-2)/2}G^{(D-3)/2}(s).  \tag{3.10}
\end{equation}
By repeating similarly the method of solution for the $D=3$ case, as
presented above, we finally obtain,

\begin{equation}
K(\mathbf{q}^{\prime \prime },t^{\prime \prime },\mathbf{q}^{\prime
},t^{\prime })=\sqrt{2}\left( -\frac 1{2\pi }\right)
^{(D-1)/2}\int\limits_d^\infty \frac{ds}{\sqrt{\cosh (s)-\cosh (d)}}\frac
\partial {\partial s}G^{(D-3)/2}.  \tag{3.11}
\end{equation}

\section{\textbf{Acknowledgements}}

The author wishes to acknowledge Prof. C. Grosche for providing him with
refs. 1 and 4. We also thank Dr. Freddy P. Zen and Dr. Alexander A. Iskandar
for the encouragements.


\begin{thebibliography}{9}
\bibitem{}  C. Grosche, \textit{Path Integral, Hyperbolic Spaces, and
Selberg Trace Formula,} (World Scientific, Singapore, 1995).

\bibitem{}  C. Grosche and F. Steiner, Phys. Lett. \textbf{A123} (1987) 319.

\bibitem{}  R. Kubo, Prog. Theor. Phys. \textbf{79} (1988) 217.

\bibitem{}  C. Grosche, Fortsch. Phys. \textbf{42 }(1994) 509.

\bibitem{}  R. P. Feynman and A. R. Hibbs, \textit{Quantum Mechanics and
Path Integrals}, (McGraw-Hill, New York, 1965).

\bibitem{}  H. Margenau and G. M. Murphy, \textit{The Mathematics of Physics
and Chemistry }(D. Van Nostrand, New York, 1965).

\bibitem{}  S. Helgason, \textit{Groups and Geometric Analysis }(Academic,
New York, 1984).

\bibitem{}  H. P. McKean, Comm. Pure and Appl. Math. \textbf{25} (1972), 225.
\end{thebibliography}
\end{document}